%% file: final.tex
\documentclass[conference]{IEEEtran}

\usepackage{amsmath,amssymb,mathrsfs,graphicx}
\usepackage{psfrag}
\usepackage{verbatim}
\input{definitions.tex}

\newtheorem{theorem}{Theorem}\def\thref#1{Theorem~\ref{#1}}

\def\figref#1{Fig.~\ref{#1}}

\def\secref#1{Section~\ref{#1}}

\normalsize

\makeatletter\def\@oddfoot{\hfill\thepage\hfill}\makeatother

\begin{document}

\title{Superposition Noisy Network Coding}

\author{ \normalsize
 \IEEEauthorblockN{Neevan~Ramalingam and Zhengdao~Wang\\}
 \small
 \IEEEauthorblockA{Dept.~of ECE, Iowa State University, Ames, IA 50011, USA.
   email: \{neevan,zhengdao\}@iastate.edu}
 }

\maketitle

\begin{abstract}
We present a superposition coding scheme for communication over a network,
which combines partial decode and forward and noisy network coding. This
hybrid scheme is termed as superposition noisy network coding. The scheme is
designed and analyzed for single relay channel, single source multicast
network and multiple source multicast network. The achievable rate region is
determined for each case. The special cases of Gaussian single relay channel
and two way relay channel are analyzed for superposition noisy network coding.
The achievable rate of the proposed scheme is higher than the existing schemes
of noisy network coding and compress-forward.
\end{abstract}

\section{Introduction}

IN an $N$-node Discrete Memoryless Network (DMN), each node transmits its
message to a set of destination nodes and acts as a relay to help transmit
messages from other nodes. It is an important network model in multi-user
information theory. This general network model includes many important
class of channels as special cases. Noiseless, erasure and deterministic
networks are few examples \cite{dago06}, \cite{rakr06}, \cite{avts09}. The DMN
also includes the relay, broadcast, interference and multiple access channels
which are the fundamental building blocks for any network. This DMN model can
also be modified to include Gaussian networks and networks with state.

The capacity of the discrete memoryless network is not known in general. The
best known upper bound is the cut-set bound \cite{coth06}. Cover and El Gamal
\cite{coga79} introduced coding schemes for the general discrete memoryless
single relay channel. The schemes introduced in brief are decode-forward,
compress-forward and superposition-forward. Superposition-forward is the
combination of decode-forward and compress-forward. Decode-forward and
compress-forward are special cases of superposition-forward. The
superposition-forward scheme achieves the optimal rate for all the special
cases where capacity is known.

Lim et al. \cite{lkg10} introduced a general lower bound for the discrete
memoryless network using the equivalent characterization of compress-forward.
This new scheme is termed ``noisy network coding". Noisy network coding
combines network coding \cite{acyl00} with the compress-forward scheme. The
key ideas used are message repetition encoding, no Wyner-Ziv \cite{wyzi76}
binning at the relay and joint decoding at the destination. This scheme
achieves a higher rate than the better known compress-forward scheme for
networks with multiple relays \cite{xiwu10-1}. The noisy network coding scheme
naturally extends to single and multiple source multicast networks.

In this paper, we improve the achievable rates of the noisy network coding
scheme by allowing the nodes to decode a part of message and use it to make a
better compressed signal to be relayed. The superposition noisy network coding
scheme combines superposition-forward with network coding. Modifications are
made to the superposition-forward scheme to make it applicable to the network
coding scenario. Specifically, the input distributions at each node are chosen
to be independent.

Superposition noisy network coding splits the message at each node into two
parts. A part of the message is required to be decoded at each relay after
every block. The other part of the message is transmitted over $b$ blocks
using repetition coding. The relay nodes use compress-forward to transmit this
message. The destination nodes decode the messages after $b$ blocks of
transmission using joint decoding. Similar to noisy network coding, our scheme
does not use Wyner-Ziv encoding at the relay, employs repetition encoding and
joint decoding. These techniques have been shown to improve the achievable
rates as in the case of network coding.

For simplicity and ease of understanding, the superposition noisy network
coding scheme is explained for a simple 3 node relay channel first. In section
II, the scheme is designed and achievable rates derived for a single relay
channel. In Section III, the scheme is further extended to single source
multicast network where there is only a single source node transmitting
information to a set of destination nodes. All other nodes can act as relays.
In Section IV, the superposition noisy network coding scheme is designed for
multiple source multicast networks. This scheme is then applied to AWGN
single and two-way relay channel to quantify performance.

\section{Superposition noisy network coding for single relay channel}

\begin{figure}
\centering
\includegraphics[width=.8\linewidth]{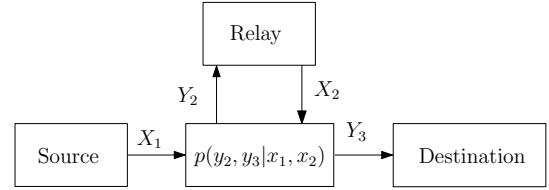}
\caption{The three node relay channel}
\label{figrelay}
\end{figure}

Consider the discrete memoryless relay channel $p(y_2,y_3|x_1,x_2)$ shown in
\figref{figrelay}. The source node is terminal 1, relay node is terminal 2,
and terminal 3 is the destination node. $x_k$ denotes the transmitted symbol at
terminal $k$. $y_k$ denotes the received symbol at terminal $k$.

The rate achieved by superposition-forward scheme \cite[Theorem 7]{coga79} for
discrete memoryless relay channel is
\begin{multline}
R_{SF}=\sup (\min \{I(X_1;Y_3,\Yh_2|X_2,U)+I(U;Y_2|X_2,V),\\
I(X_1,X_2;Y_3)-I(\Yh_2;Y_2|U,X_1,X_2,Y_3)\} ) \label{eq.th7}
\end{multline}
where the supremum is over all joint probability distributions of the form
\begin{equation}\label{eq.p7}
\begin{split}
&p(u,v,x_1,x_2,\yh_2,y_3,y_2) = \\
& p(v)p(u|v)p(x_1|u) p(x_2|v)p(y_2,y_3|x_1,x_2)p(\yh_2|x_2,y_2,u)
\end{split}
\end{equation}
subject to the constraint
\begin{equation}\label{eq.th7con}
I(X_2;Y_3|V) \ge I(\Yh_2;Y_2|X_2,Y_3,U).
\end{equation}

To facilitate network coding we will restrict the superposition-forward
strategy such that the auxiliary random variables $U$ and $V$ are generated
independent of each other. This will lead to a rate loss compared to original
scheme. Nevertheless, the rates achieved would be higher than the
compress-forward or noisy network coding scheme.

The rate achieved by superposition noisy network coding scheme for a single
relay channel is stated in \thref{th.1}.

\medskip
\begin{theorem}{}\label{th.1}
For any discrete memoryless relay channel, the rate $\sup_P R'+R''$ is
achievable, where
\begin{align}
R' &< \min \{I(U_1;Y_2|X_2),\, I(U_1,V_2;Y_3) \}, \\
R'' &< \min \{I(X_1; \Yh_2,Y_3|X_2,U_1),\, I(X_1,X_2;Y_3|U_1,V_2) -\notag \\
&\qquad\qquad I(\Yh_2;Y_2|U_1,X_1,X_2,Y_3)\},
\end{align}
and the supremum is taken over all joint probability distributions of the form
\begin{align}
p(u_1,v_2,x_1,x_2,y_2,y_3,\yh_2) = p(u_1)p(v_2)p(x_1|u_1)\cdot \notag \\
p(x_2|v_2)p(y_2,y_3|x_1,x_2)p(\yh_2|x_2,y_2,u_1)
\end{align}
\end{theorem}

\begin{IEEEproof}
The message $m'\in [1:2^{nR'}]$ is transmitted over every block $j$ and the
message $m''\in [1:2^{nbR''}]$ is transmitted over $b$ blocks of transmission.
The source node transmits $\xv_{1j}(m''|m_j')$ for each block $j\in[1:b]$.
After block $j$, the relay decodes the message $m_j'$ and maps it to a
$\vv_{2, j+1j}(m_j')$ codeword. It also finds a ``compressed'' version
$\yvh_{2j}(l_j|l_{j-1}, m_j', m_{j-1}')$ of the relay output $\yv_{2j}$
conditioned on $\xv_{2j}$ and $\uv_{1j}$. The relay transmits a
codeword $\xv_{2, j+1}(l_j|m_j')$
in the next block. After $b$ blocks of transmission, the decoder finds the
correct message $m''\in [1:2^{nbR}]$ using $(\yv_{31}, \ldots, \yv_{3b})$ and
joint decoding for each of the $b$ blocks simultaneously. The decoder has
decoded all the messages $m'_j$ by sliding window decoding for each
block $j \in [1:b]$. The details are as follows.

\medskip

\begin{table*}[t]
\begin{center}
\tiny
\begin{tabular}{c|cccccc}
\hline \textrm{Block} & 1 & 2 & 3 & $\cdots$ & $b-1$ & $b$\\ \hline%
$U_1$ & $\uv_{11}(m_1')$ &
$\uv_{12}(m_2')$ & $\uv_{13}(m_3')$ & $\ldots$ & $\uv_{1, b-1}(m_{b-1}')$ &
$\uv_{1b}(m_b')$\\
$V_2$ & $\vv_{21}(1)$ & $\vv_{22}(m_1')$ & $\vv_{23}(m_2')$ & $\ldots$ &
$\vv_{2, b-1}(m_{b-2}')$ & $\vv_{2b}(m_{b-1}')$\\
$X_1$ & $\xv_{11}(m''|m_1')$ & $\xv_{12}(m''|m_2')$ &
$\xv_{13}(m''|m_3')$ & $\ldots$ & $\xv_{1, b-1}(m''|m_{b-1}')$ &
$\xv_{1b}(m''|m_b')$\\ $\Yh_{2}$ & $\yvh_{21}(l_1|1,1,m_1')$ &
$\yvh_{22}(l_2|l_1,m_1',m_2')$ & $\yvh_{23}(l_3|l_2,m_2',m_3')$ & $\ldots$ &
$\yvh_{2, b-1}(l_{b-1}|l_{b-2},m_{b-2}',m_{b-1}')$ &
$\yvh_{2b}(l_b|l_{b-1},m_{b-1}',m_b')$ \\ $X_2$ & $\xv_{21}(1|1)$ &
$\xv_{22}(l_1|m_1')$ & $\xv_{23}(l_2|m_2')$ & $\ldots$ & $\xv_{2,
b-1}(l_{b-2}|m_{b-2}')$ & $\xv_{2b}(l_{b-1}|m_{b-1}')$\\ $Y_3$ & $\emptyset$ &
$\mh_1'$ & $\mh_2'$ & $\ldots$ & $\mh_{b-2}'$ & $\mh_{b-1}'$,$\mh''$\\ \hline
\end{tabular}
\end{center}
\caption{Superposition noisy network coding for the relay channel.}
\label{tb:coding}
\end{table*}

\noindent\emph{Codebook generation:} Fix
$p(u_1)p(x_1|u_1)p(v_2)p(x_2|v_2)\cdot$ $p(\yh_2|y_2,x_2,u_1)$.

\begin{enumerate}
\item For each $j \in [1:b]$, randomly and independently generate $2^{nR'}$
sequences $\uv_{1j}(m'_j), $ $m' \in [1:2^{nR'}],$ each according to
$\prod_{i=1}^n p_{U_1}(u_{1,(j-1)n+i})$.

\item For each $\uv_{1j}(m'_j),$ randomly and independently generate
$2^{nbR''}$ sequences
$\xv_{1j}(m''|m_j'),$ $m'' \in [1: 2^{nbR''}],$
each according to
$\prod_{i=1}^n p_{X_1|U_1}(x_{1, (j-1)n+i}|u_{1,(j-1)n+i}(m'_j))$.

\item Similarly, randomly and independently generate $2^{nR'}$ sequences
$\vv_{2j}(m'_{j-1}), $ $m_{j-1}' \in [1:2^{nR'}],$ each according to
$\prod_{i=1}^n p_{V_2}(v_{2,(j-1)n+i})$.

\item For each $\vv_{2j}(m'_{j-1}),$ randomly and independently
generate $2^{n\Rh_2}$ sequences
$\xv_{2j}(l_{j-1}|m_{j-1}'),$ $l_{j-1} \in
[1:2^{n\Rh_2}],m_{j-1}' \in [1:2^{nR'}],$ each according to \\ $\prod_{i=1}^n
p_{X_2|V_2}(x_{2, (j-1)n+i}|v_{2, (j-1)n+i}(m'_{j-1}))$.

\item For each $\xv_{2j}(l_{j-1}|m'_{j-1}),$ $l_{j-1} \in [1:2^{n\Rh_2}]$ and
$\uv_{1j}(m_j'),$ $m_j', m_{j-1}' \in [1:2^{nR'}],$
randomly and independently generate $2^{n\Rh_2}$ sequences
$\yvh_{2j}(l_j|l_{j-1},m_{j-1}',m_j'),$ $l_j \in [1: 2^{n\Rh_2}]$, each
according to $\prod_{i=1}^n p_{\Yh_2|X_2,U_1}
(\yh_{2, (j-1)n+i}|x_{2, (j-1)n+i}(l_{j-1},m_{j-1}'),$
$u_{1, (j-1)n+i}(m_j'))$.
\end{enumerate}

This defines the codebook
\begin{align}
 \Cc_j &= \bigl\{\uv_{1j}(m_j'), \vv_{1j}(m_{j-1}'), \xv_{1j}(m''|m_j'),
  \xv_{2j}(l_{j-1}|m_{j-1}'), \notag \\
  & \qquad \yvh_{2j}(l_j|l_{j-1},m_{j-1}', m_j') : m_j',m_{j-1}'
\in [1:2^{nR'}],
  \notag \\
  & \qquad  m'' \in [1:2^{nbR''}],
   l_j, l_{j-1} \in [1:2^{n\Rh_2}]\bigr\}
\end{align}
for $j \in [1:b]$.

Encoding and decoding are explained with the help of Table~\ref{tb:coding}.

\noindent\emph{Encoding:} Let $m_j'$ be the message to be sent in block $j$
and $m''$ be the message to be sent over $b$ blocks. The relay, upon receiving
$\yv_{2j}$ at the end of block $j \in [1:b]$, finds an index $m_j'$ such that
\[
(\uv_{1j}(m_j'),\yv_{2j}, \xv_{2j}(l_{j-1}|\mh_{j-1}')) \in
\aepvar,
\]
and then finds an index $l_j$ such that
\begin{align}
(\uv_{1j}(\mh_{j-1}'), \yvh_{2j}(l_j|l_{j-1},&\mh_j',\mh_{j-1}'),
  \yv_{2j},\notag  \\
  &\xv_{2j}(l_{j-1}|\mh_{j-1}')) \in \aepvar, \notag
\end{align}
where $l_{0}=1$ by convention. If there is more than one such index, choose
one of them at random. If there is no such index, choose an arbitrary index at
random from $[1:2^{n\Rh_2}]$. The codeword pair $(\xv_{1j}(m''|m_j'),
\xv_{2j}(l_{j-1}|m_{j-1}'))$ is transmitted in block $j\in [1:b]$.

\medskip \noindent\emph{Decoding:} Let $\e > \e'$. After block $j$, the
decoder uses $\yv_{3(j-1)}$ and $\yv_{3j}$ to find a unique message
$\mh_{j-1}' \in [1:2^{nR'}]$. The unique message satisfies the following two
conditions simultaneously
\begin{eqnarray}
(\uv_{1j}(m_{j-1}'),\vv_{2j}(\mh_{j-2}'), \yv_{3{j-1}}) &\in \aep \nonumber\\
(\vv_{2j}(m_{j-1}'), \yv_{3j}) &\in \aep \nonumber
\end{eqnarray}

\noindent At the end of block $b$, after decoding all the messages $m_j'$,
$j\in [1:(b-1)]$ the decoder finds a unique message
$\mh'' \in [1:2^{nbR''}]$ such that
\begin{align*}
(\uv_{1j}(\mh_j'), &\vv_{2j}(\mh_{j-1}'), \yvh_{2j}(l_j| l_{j-1},
   \mh_{j-1}', \mh_j'), \xv_{1j}(m_j''|\mh_{j-1}'), \\
  & \xv_{2j}(l_{j-1}|\mh_{j-1}'), \yv_{3j}) \in \aep,
    \text{ for all } j \in [1:b]
\end{align*}
for some $l_1, l_2, \ldots, l_b$. If there is none or more than one such
message, it declares an error.

\medskip \noindent\emph{Analysis of the probability of error:} Let $M_j'$,
denote the messages sent at the source node for $j \in [1:(b-1)]$, $M''$ be the
message sent at the source node over $b$ blocks and $L_j$ denote the indices
chosen by the relay at block $j\in [1:b]$. Define
\begin{align*}
& \Ec_{m'(0)} := \bigcup_{j=1}^b \bigl\{ (\Uv_{1j}(m_j'),\Yv_{2j},
  \Xv_{2j}(l_{j-1}|m_{j-1}')) \not \in \aepvar \bigr\},\\
& \Ec_{m'(1)} :=\bigl\{(\uv_{1,j-1}(m_{j-1}'),\vv_{2,j-1}(\mh_{j-2}'),
  \yv_{3{j-1}}) \in \aep,\, \\
  & \qquad j\in[1:b] \bigr\} \bigcup
    \bigl\{(\vv_{2j}(m_{j-1}'), \yv_{3j}) \in \aep,\, j\in[1:b] \bigr\}. \\
\end{align*}

To bound the probability of error in decoding message $m_{j-1}'$,
assume without loss of generality that
$M_{j-2}'=M_{j-1}'=1$. The probability of error is upper bounded by

\begin{align*}
\P(\Ec)
& \leq \P(\Ec_{m'(0)}) + \P(\Ec_{m'(0)}^{c} \cap \Ec_{m'(1)}^c \cap M_{j-1}' =1)
\\
& \quad + \P(\Ec_{m'(0)}^{c} \cap \Ec_{m'(1)} \cap M_{j-1}' \neq 1).
\end{align*}

By the conditional typicality lemma \cite{gaki10}, $\P(\Ec_{m'(0)}) \to 0$ as
$n \to \infty$ if $R' < I(U_1;Y_2|X_2)$ for sufficiently large $n$, and
$\P(\Ec_{m'(0)}^{c} \cap \Ec_{m'(1)}^c \cap M_{j-1}' =1) \to 0$ as $n \to
\infty$. Since the codebooks are generated independently for each block, the
two events of $\Ec_{m'(1)}$ are independent. Thus by the law of large numbers
and joint typicality lemma \cite{gaki10} $\P(\Ec_{m'(0)}^{c} \cap \Ec_{m'(1)}
\cap M_{j-1}' \neq 1) \to 0$ as $n \to \infty$ if
\[
R' < I(U_1;Y_3|V_2) + I(V_2;Y_3) = I(U_1,V_2;Y_3)
\]
and $n$ is sufficiently large. So the message $m_j'$ can be decoded correctly
at the destination provided

\begin{align*}
R' < \min\{I(U_1;Y_2|X_2),I(U_1,V_2;Y_3)\}
\end{align*}

After decoding the messages $m_j'$ for $j\in [1:(b-1)]$, the destination
decodes the message $M''$ after $b$ blocks. The probability of error analysis
for message $M''$ is similar to the noisy network coding scheme \cite{lkg10},
given the partial information of the messages $m_j'$. It can be shown that
when
\begin{align}
R'' & < \min \{I(X_1; \Yh_2,Y_3|X_2,U_1),\, I(X_1,X_2;Y_3|U_1,V_2) - \notag  \\
  & \quad I(\Yh_2;Y_2|U_1,X_1,X_2,Y_3)\} - \delta(\e) - \delta(\e'), \notag
\end{align}
the probability of error of detecting $M''$ can be made arbitrarily small. The
probability of error analysis is omitted here due to limited space.
\end{IEEEproof}

\section{Superposition Noisy Network Coding for Multicast Networks}
\label{sec.multicast}

We now describe the superposition noisy network coding scheme for
single-source discrete memoryless networks with multicast (DMN-MC)
$p(y_2,\ldots,y_N|x^N)$, where terminal 1 is the source node. We assume that
there is no feedback to terminal~1. Source terminal 1 splits the message in
two parts $m'$ and $m''$ and transmits using superposition forwarding. The
message $m'$ is transmitted in the same fashion decode-forward is extended to
multicast relay networks \cite{kgg05}, \cite{xiku04}. The scheme is modified
to make the input distributions at each node independent of each other. After
decoding the partial information $m'$, the message $m''$ is decoded using
noisy network coding \cite{lkg10} given the partial information.

\medskip
\begin{theorem}{}
For a discrete memoryless multicast network $p(y_2,\ldots,y_N|x^N)$, the rate
$R'+R''$ is achievable, where
\begin{align*}
R' &< \min_{k} I(V^{k-1};Y_k|X_k,V_k^N) \\
R'' &<  \min_{\Sc} \left( I(X(\Sc);\Yh(\Sc^c), Y_k|X(\Sc^c), V) \right. - \\
  & \qquad \qquad \left. I(\Yh(\Sc); Y(\Sc) | X^N, \Yh(\Sc^c), Y_k, V_{k-1}^N)
    \right)
\end{align*}
and $k \in \Dc$ the set of destination nodes. The minimum is over all possible
cut-sets for node $k$. The random variables satisfy a joint pmf of the form
$p(v_1)p(x_1|v_1) \prod_{k=2}^N p(v_k)p(x_k|v_k)p(\yh_k|y_k,x_k,v_k^N)$.
\end{theorem}

\begin{IEEEproof}[Sketch of Proof]
The encoding and decoding process is similar to superposition noisy network
coding for single relay channel. The relay nodes use an extension of
decode-forward to multicast networks. The partial message is decoded at each
of the nodes $\{ 1:(k-1) \}$, and coherently transmitted to node $k$. The node
$k$ waits for $k-1$ transmissions to decode the partial information. After
decoding the partial message, the remaining message is decoded using noisy
network coding. Due to space limit, we omit details of the encoding and
decoding processes, and the error probability analysis.
\end{IEEEproof}

\section{Superposition Noisy Network Coding for Multiple Source Multicast
Networks}

The superposition noisy network coding scheme can also be generalized to an
$N$ node discrete memoryless multiple source multicast network $p(y^N|x^N)$,
\cite{lkg10}. In the general setup each node sends its independent message to
a set of destination nodes while acting as relays for messages from other
sources.

We make a general assumption to make the application of superposition noisy
network coding easier. The source nodes are restricted not to act as relays.
Two-way relay channel and interference relay channel are two examples where
such an assumption holds. With this assumption, the channel model is now
similar to single source multicast network with a replacement of the source
node with many independent nodes. The partial information is transmitted the
same way decode-forward is extended for the single source multicast network in
\secref{sec.multicast}. The relay decodes the message from all the sources
using an $m$-user multiple access channel \cite{coth06}. After decoding the
partial information from the source nodes, the relay uses binning to transmit
the decoded information \cite{xie07}. Further relays and destination nodes
decode the message in the same multiple access fashion. The relays that have
decoded the messages act as source nodes and coherently transmit the partial
information. The remaining message is superimposed and decoded using noisy
network coding. The following theorem provides an achievable rate for this
network, using superposition noisy network coding.

\begin{theorem}{}\label{thm:multi}
For an $N$ node discrete memoryless multiple source multicast network with
$k_0$ source nodes, the following rate is achievable using superposition noisy
network coding
\begin{align}
R'(\Sc) &< \min_{\Sc} I(V(\Sc);Y_k|X_k,V(\Sc^c)) \label{eq.RS} \\
R''(\Sc) &<  \min_{\Sc} \bigl( I(X(\Sc);\Yh(\Sc^c), Y_N|X(\Sc^c), V_1^N)
  \notag \\
 & \qquad - I(\Yh(\Sc); Y(\Sc) | X^N, \Yh(\Sc^c), Y_N, V_{k-1}^N) \bigr)
 \label{eq.RS2}
\end{align}
where the random variables are jointly distributed according to
$\prod_{k=1}^{k_0} p(v_k)p(x_k|v_k) \prod_{k=k_0+1}^N
p(v_k)p(x_k|v_k)p(\yh_k|y_k,x_k,u_1)$, and the maximum is over all possible
cut-sets $\Sc$.
\end{theorem}
\begin{IEEEproof}[Sketch of Proof] \\
\noindent\emph{Codebook generation:} Fix \\ $\prod_{k=1}^{k_0}
p(v_k)p(x_k|v_k) \prod_{k=k_0+1}^N p(v_k)p(x_k|v_k)p(\yh_k|y_k,x_k,u_k^N)$.
\begin{enumerate}
\item For each $j \in [1:b]$ and $k \in [1:k_0]$, randomly and independently
generate $2^{nR_k'}$ sequences $\vv_{k,j}(m_k')$, $m_k' \in [1:2^{nR_k'}]$,
each according to $\prod_{i=1}^n p_{V_k}(v_{k,(j-1)n+i})$.

\item For each $\vv_{k,j}(m_k')$, $j \in [1:b]$ and $k \in [1:k_0]$, randomly
and conditionally independently generate $2^{nbR_k''}$ sequences $\xv_{k,
j}(m_k''|m_k'),$ such that $m_k'' \in [1:2^{nbR_k''}]$, $m_k' \in
[1:2^{nbR_k'}]$. The sequences are generated independently according to the
distribution\\ $\prod_{i=1}^n p_{X_k|V_k}(x_{k, (j-1)n+i}|v_{k,
(j-1)n+i}(m_k'))$

\item For all nodes $k \in [k_0+1 : N]$ randomly and independently generate
$2^{n\tilde{R}_k'}$ codewords $\vv_{k,j}(\kappa(m_1^{'k_0}))$. The rate
$\tilde{R}_k'$ is chosen such that
\[
\tilde{R}_k' \geq \max_{d \in \Dc} I(V_k;Y_d|V_1^{k_0})
\]
The maximum is over $\Dc$ the set of all destination nodes.
$\kappa(m_1^{'k_0})$ is the bin index of the messages $m_1^{'k_0}$.

\item For each $\vv_{k,j}(\kappa(m_1^{'k_0}))$ and $k \in [k_0+1 : N]$,
randomly and independently generate $2^{n\Rh_k}$ sequences $\xv_{k, j}(l_{k,
j-1}|\kappa(m_1^{'k_0})),$ such that $m_k' \in [1:2^{nR_k'}]$, $l_{k, j-1} \in
[1:2^{n\Rh_k}]$, each according to the probability distribution\\
$\prod_{i=1}^n p_{X_k|V_k}(x_{k, (j-1)n+i}|v_{k,
(j-1)n+i}(\kappa(m_1^{'k_0})))$.

\item For each node $k \in [k_0+1:N]$ and each $\xv_{k, j}(l_{k,
j-1}|\kappa(m_1^{'k_0}))$ $\vv_{kj}(\kappa(m_1^{'k_0})),\ldots,
\vv_{Nj}(\kappa(m_1^{'k_0}))$, such that $m_k'' \in [1:2^{nbR_k''}]$, $m_k'\in
[1:2^{nR_k'}],$ $l_{k, j-1} \in [1:2^{n\Rh_k}],$ randomly and conditionally
independently generate $2^{n\Rh_k}$ sequences $\yvh_{kj}(l_{kj}|m_k'', l_{k,
j-1}, \kappa(m_1^{'k_0})),$ $l_{kj} \in [1: 2^{n\Rh_k}],$ each according to \\
$\prod_{i=1}^n p_{\Yh_k |X_k, V_k} (\yh_{k,(j-1)n+i}|x_{k,
(j-1)n+i}(m_k''|m_k'),\\
v_{k, (j-1)n+i}(\kappa(m_1^{'k_0})))$.
\end{enumerate}

This defines the codebook
\begin{eqnarray}
\Cc_j &=& \bigl\{\vv_{k,j}(m_k'), \xv_{k, j}(m_k''|m_k'), k \in [1:k_0], \notag \\
&& \vv_{k,j}(\kappa(m_1^{'k_0})), \xv_{k, j}(l_{k,
j-1}|\kappa(m_1^{'k_0})), \notag \\
&& \qquad \yvh_{kj}(l_{kj}|m_k'', l_{k, j-1}, \kappa(m_1^{'k_0})), k \in [k_0+1:N] \notag\\
&&: m_k' \in [1:2^{nR_k'}], m_k'' \in [1:2^{nbR_k''}], \notag \\
&& l_{kj}, l_{k,j-1} \in [1:2^{n\Rh_k}] \bigr\} \nonumber
\end{eqnarray}
for $j \in [1:b]$.

\medskip

\noindent\emph{Encoding:} Let $(m_1', \ldots, m_{k_0}',m_1'', \ldots,
m_{k_0}'')$ be the messages to be sent. Each relay node $k\in [k_0+1:N]$, upon
receiving $\yv_{kj}$ at the end of block $j\in [1:b]$, decode the messages
$m_1^{'k_0}$ as shown in the decoding step. After finding $m_1^{'k_0}$, the
node finds an index $l_{kj}$ such that
\begin{align*}
& (\yvh_{kj}(l_{kj}|m_k'', l_{k, j-1}, \kappa(m_1^{'k_0})),\yv_{kj},
  \xv_{k, j}(l_{k, j-1}|\kappa(m_1^{'k_0})), \notag \\
& \qquad \qquad  \vv_{k,j}(\kappa(m_1^{'k_0}))) \in \aepvar,
\end{align*}
where $l_{k0}=1$, $k\in[k_0+1:N]$, by convention. If there is more than one
such index, choose one of them at random. If there is no such index, choose an
arbitrary index at random from $[1:2^{n\Rh_k}]$. Then each node $k\in
[k_0+1:N]$ transmits the codeword $\xv_{k, j}(l_{k, j-1}|\kappa(m_1^{'k_0}))$
in block $j \in [1:b]$.

\medskip

\noindent\emph{Decoding:} Let $\e > \e'$. After each block, the decoder $d\in
\Dc$ decodes the messages $m_1^{'k_0}$. The messages are decoded as a $k_0$
user multiple access channel. The probability of error of decoding the
messages $m_1^{'k_0}$ can be arbitrarily small if the \eqref{eq.RS} is
satisfied \cite{coth06}. After $b$ blocks, the decoder $d\in \Dc$ finds a
unique index tuple $(\mh_{1d}'',\ldots,\mh_{k_0d}'')$, where $
\mh_{kd}''\in[1:2^{nbR_k''}]$, such that there exist some $(\lh_{1j}, \ldots,
\lh_{Nj})$, $\lh_{kj}\in[1:2^{n\Rh_k}]$, and $j\in[1:b]$, satisfying
\begin{align*}
&(\vv_{1,j}(m_1'), \ldots, \vv_{k_0,j}(m_{k_0}'),
\vv_{{k_0+1},j}(\kappa(m_1^{'k_0})), \ldots, \\
& \quad \vv_{Nj}(\kappa(m_1^{'k_0})), \xv_{1, j}(m_1''|m_1'), \ldots, \xv_{k_0, j}(m_{k_0}''|m_{k_0}'), \\
& \quad \xv_{(k_0+1), j}(l_{(k_0+1), j-1}|\kappa(m_1^{'k_0})), \ldots, \xv_{N, j}(l_{N, j-1}|\kappa(m_1^{'k_0})),
\\
& \quad \yvh_{(k_0+1),j}(l_{(k_0+1),j}|l_{k_0+1, j-1}, \kappa(m_1^{'k_0})), \ldots \\
& \quad \qquad \qquad \yvh_{Nj}(l_{Nj}|l_{k, j-1}, \kappa(m_1^{'k_0})), \yv_{Nj})\in\aep
\end{align*}
for all $j\in[1:b]$, given that the messages $m_1^{'k_0}$ have been decoded
correctly. The probability of error goes to 0 as $n\to\infty$ if \eqref{eq.RS2}
is satisfied. The detailed analysis is similar to that in \cite{lkg10} and is
omitted here.
\end{IEEEproof}

\section{Numerical Results}

In this section, we apply the superposition noisy network coding scheme to
additive white Gaussian noise (AWGN) three-node relay channel and the two-way
relay channel. We compare the achievable rates to the existing schemes of
noisy network coding, compress-forward and the cut-set upper bound.

Consider a Gaussian relay channel model \cite{kgg05}
\begin{eqnarray}
Y_2 &=& a X_1 + Z_1 \\ Y_3 &=& X_1+b X_2 + Z_2
\end{eqnarray}
where the noise terms $Z_1$ and $Z_2$ are uncorrelated zero mean Gaussian
random variables with variances $N_1$ and $N_2$ respectively, and $a$ and $b$
are the channel gain constants. The power constraints at the transmitters are
\(
\frac{1}{n} \sum_{i=1}^n x_{1i}^2(k) \le P_1, \quad \forall k \in \Mc
\), and
\(\frac{1}{n} \sum_{i=1}^n x_{2i}^2  \le P_2,
  \quad \forall y_2^n \in {\mathcal R}^n
\).

\begin{figure}
\centering
\includegraphics[width=\linewidth]{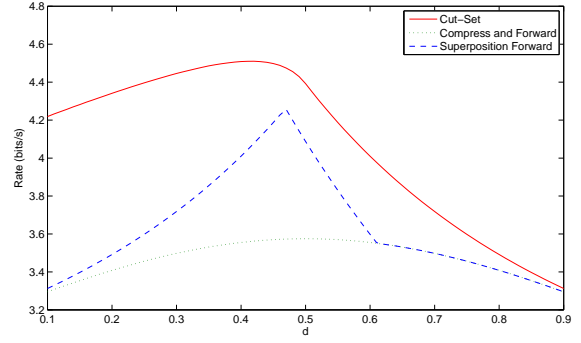}
\caption{Achievable rates for an AWGN single relay channel}
\label{figsnnc}
\end{figure}

All the terminals are aligned in a line. The source and destination are at
unit distance. The relay is at distance $d$ from the source and distance $1-d$
from the destination. We assume $a=1/d$ and $b=1/(1-d)$. \figref{figsnnc}
plots the rates achieved by superposition noisy network coding for $P1=P2=5$.
They are compared to those achieved by noisy network coding, compress-forward
and the cut-set bound. Noisy network coding achieves the same rate as
compress-forward scheme for a single relay channel.

It is observed that the superposition noisy network coding scheme has an
advantage over the noisy network coding scheme when the relay is close to the
source. This advantage arises due to a strong source-relay link.

\subsection{Two-Way Relay Channel}

The two-way relay channel was first introduced by Shannon \cite{shan61}. The
two-way relay channel is a fundamental building block for multi-user
information theory. Rankov et al. \cite{rawi06} derived the achievable rates
for the two-way relay channel using the schemes decode-forward and
compress-forward. The rates achieved by superposition noisy network coding is
derived for the two way relay channel and compared to the existing rates.

Consider the AWGN two-way relay channel \cite{rawi06}
\begin{align}
Y_1&=g_{21}X_2+g_{31}X_3+Z_1, \nonumber\\
Y_2&=g_{12}X_1+g_{32}X_3+Z_2, \label{eq:awgn-twrc}\\
Y_3&=g_{13}X_1+g_{23}X_2+Z_3, \nonumber
\end{align}
where the channel gains are $g_{12} = g_{21} = 1$, $g_{13} = g_{31} =
d^{-\gamma/2}$ and $g_{23} = g_{32} = (1-d)^{-\gamma/2}$, and $d \in [0,1]$ is
the location of the relay node between nodes 1 and 2 (which are unit distance
apart). Source nodes $1$ and $2$ wish to exchange messages reliably with the
help of relay node $3$. Specializing the \thref{thm:multi} to the two-way
relay channel gives the inner bound that consists of all rate pairs
$(R_1,R_2)$ such that
\begin{eqnarray*}
R_1' &\le& \min\{I(U_1;Y_2|U_2, V_3, X_3), \\ && \qquad \qquad I(U_1,
V_3;Y_2|U_2,X_2)\}\\ R_2' &\le& \min\{I(U_2;Y_1|U_1, V_3, X_3),\\ && \qquad
\qquad I(U_2, V_3;Y_1|U_1,X_1)\}\\ R_1'+R_2' &\le&I(U_1,U_2;Y_3|V_3,X_3) \\
R_1'' &\le& \min\{I(X_1; Y_2, \Yh_3|X_2, X_3, U_1, U_2),\,\\ && \qquad I(X_1,
X_3; Y_2|X_2, U_1, V_3)- \\ && \qquad
 \qquad I(Y_3; \Yh_3|X_1, X_2, X_3, Y_2, U_1, U_2)\}\\ R_2'' &\le& \min\{I(X_2; Y_1, \Yh_3|X_1, X_3, U_1, U_2),\,
\\ && \qquad I(X_2, X_3; Y_1|X_1, U_2, V_3)- \\ && \qquad \qquad I(Y_3;
\Yh_3|X_1, X_2, X_3, Y_1, U_1, U_2)\}
\end{eqnarray*}
for some $p(q)p(u_1)p(u_2)p(v_3)p(x_1|u_1,q)p(x_2|u_2,q)p(x_3|v_3,q)\\p(\yh_3|
y_3, x_3, q)$.

\begin{figure}
\centering
\includegraphics[width=\linewidth]{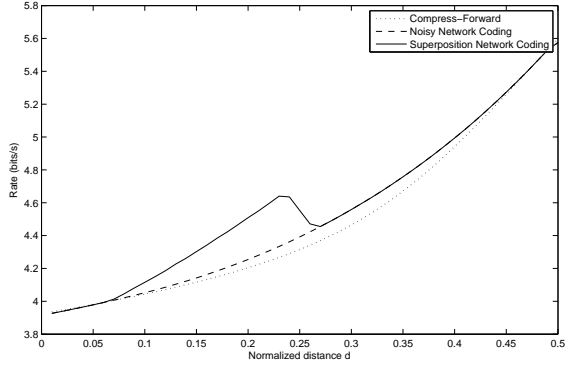}
\caption{Achievable rates for an AWGN two-way relay channel}
\label{figtwrc}
\end{figure}

\figref{figtwrc} compares the achievable rates of the schemes derived as a
function of relay distance. The power constraints at the nodes are $P_1 = P_2
= P_3 = 10$. It is observed that superposition noisy network coding provides
higher rates than both compress-forward and noisy network coding. Noisy network
coding is a special case of superposition noisy network coding scheme. The
superposition scheme performs better when the relay is close to either of the
sources and decoding partial information is advantageous to the sum rate.

\section{Conclusions}

The noisy network coding for discrete memoryless channel is improved by
superimposing partial decode and forward of the messages. The encoding and
decoding strategies are first derived for the superposition noisy network
coding in a three-node relay channel. The rates achieved by superposition
noisy network coding is higher than the rates achieved by noisy network
coding, when the channel from the source to the relay nodes are strong. We
then derive the superposition noisy network coding scheme for both
single-source and multiple-source multicast networks. We specialized the
result to a two-way relay channel. For Gaussian three-node and two-way relay
channels, it is numerically observed that the superposition noisy network
coding scheme provides higher rates than noisy network coding or
compress-forward.

\bibliographystyle{IEEEtran}
\bibliography{refs}
\end{document}

%% file: definitions.tex
\newcommand{\Cc}{\mathcal{C}}
\newcommand{\Dc}{\mathcal{D}}
\newcommand{\Ec}{\mathcal{E}}
\newcommand{\Sc}{\mathcal{S}}
\newcommand{\Mc}{\mathcal{M}}

\newcommand{\Xv}{\mathbf{X}}
\newcommand{\Yv}{\mathbf{Y}}
\newcommand{\Uv}{\mathbf{U}}

\newcommand{\xv}{\mathbf{x}}
\newcommand{\yv}{\mathbf{y}}
\newcommand{\uv}{\mathbf{u}}
\newcommand{\vv}{\mathbf{v}}

\newcommand{\aep}{{\mathcal{T}_{\epsilon}^{(n)}}}
\newcommand{\aepvar}{{\mathcal{T}_{\epsilon'}^{(n)}}}

\let\P\relax
\DeclareMathOperator\P{\sf P}

\newcommand{\Rh}{{\hat{R}}}

\newcommand{\Yh}{{\hat{Y}}}
\newcommand{\lh}{{\hat{l}}}
\newcommand{\mh}{{\hat{m}}}

\newcommand{\yh}{{\hat{y}}}

\newcommand{\yvh}{\hat{\mathbf{y}}}

\def\e{\epsilon}